\documentclass[aps,prl,twocolumn,amsmath,amssymb,nofootinbib,superscriptaddress,floatfix]{revtex4}

\usepackage{amsmath}
\usepackage{amssymb}
\usepackage{amsthm}
\usepackage[pdftex]{color} 
\usepackage{graphicx}
\usepackage{dcolumn} 
\usepackage{bm} 
\usepackage[driverfallback=dvipdfm]{hyperref} 

\usepackage{longtable}
\usepackage{ulem}   
\newcommand{\ket}[1]{{|\,{#1}\,\rangle}}
\newcommand{\bra}[1]{{\langle\,{#1}\,|}}

\begin{document}

\author{Luiz H. Santos}
\affiliation
{
Department of Physics and Institute for Condensed Matter Theory,
University of Illinois at Urbana-Champaign, 1110 West Green Street, Urbana, Illinois, 61801-3080, USA
}

\author{Taylor L. Hughes}
\affiliation
{
Department of Physics and Institute for Condensed Matter Theory,
University of Illinois at Urbana-Champaign, 1110 West Green Street, Urbana, Illinois, 61801-3080, USA
}

\title{
     Parafermionic wires at the interface of chiral topological states
      }

\begin{abstract}
We explore a scenario where local interactions form one-dimensional gapped interfaces 
between a pair of distinct chiral two-dimensional topological states -- referred to as phases 1 and 2 -- 
such that each gapped region terminates at a domain wall separating the chiral gapless edge states of these phases.
We show that this type of T-junction supports point-like fractionalized excitations obeying parafermion statistics,
thus implying that the one-dimensional gapped interface forms an effective topological parafermionic wire
possessing a non-trivial ground state degeneracy. %
The physical properties of the anyon condensate that gives rise to the gapped interface are investigated. 
Remarkably, this condensate causes the gapped interface to behave as a type of
anyon ``Andreev reflector" in the bulk, whereby anyons from one phase,
upon hitting the interface, can be transformed into a combination of
reflected anyons and outgoing anyons from the other phase. Thus, we conclude that while different topological orders can be connected via gapped interfaces, the interfaces are themselves topological.

\end{abstract}

\date{\today}

\maketitle



\section{Introduction}
\label{sec: introduction}

Topological phases (TPs) of matter in two dimensions (2D) are often characterized by
a ``bulk-boundary" correspondence. Bulk properties
such as a topological band structure,  quasiparticles 
exhibiting fractional statistics, or topological ground state
degeneracy on manifolds with non-zero genus, go hand in hand with  
an associated set of boundary/interface states where a 
TP 
meets a different 
one such as the vacuum.~\cite{Wen-Book}

TPs
appear in 
two general classes: symmetry protected~\cite{Hasan-2010,Moore-2010,Qi-2011,Schnyder-2008,Qi-2008,Kitaev-2009,Pollmann-2010,Schuch-2011,Chen-2013,Levin-2012,Lu-2012,Vishwanath-2013,C-Wang-2014,Bi-2013},
or those that have ``intrinsic" topological order~\cite{Wen-1995}. 
There are several important distinctions between these classes, e.g., differing constraints on the ability to open a gap in the edge state spectrum. 
For the first class, gapped boundaries can exist when the symmetry is
broken explicitly or spontaneously. In the latter, interface states with non-vanishing chirality cannot be completely gapped, and, surprisingly, even in the absence of any symmetries,  some interfaces with \emph{vanishing} chirality cannot be completely gapped either.\cite{Levin-2013}  This observation may directly impact experiment since such an ungappable edge may exist in the $\nu=2/3$ fractional quantum Hall effect, or at the interface
between two fractional quantum Hall states 
with, e.g., filling factors $\nu = 1/3$ and $\nu = 1/5.$ The latter interface cannot be gapped by any local interaction, essentially due to the completely incompatible bulk 
properties of the two 
TPs.

\begin{figure}[h!]
\includegraphics[width=0.5\textwidth]{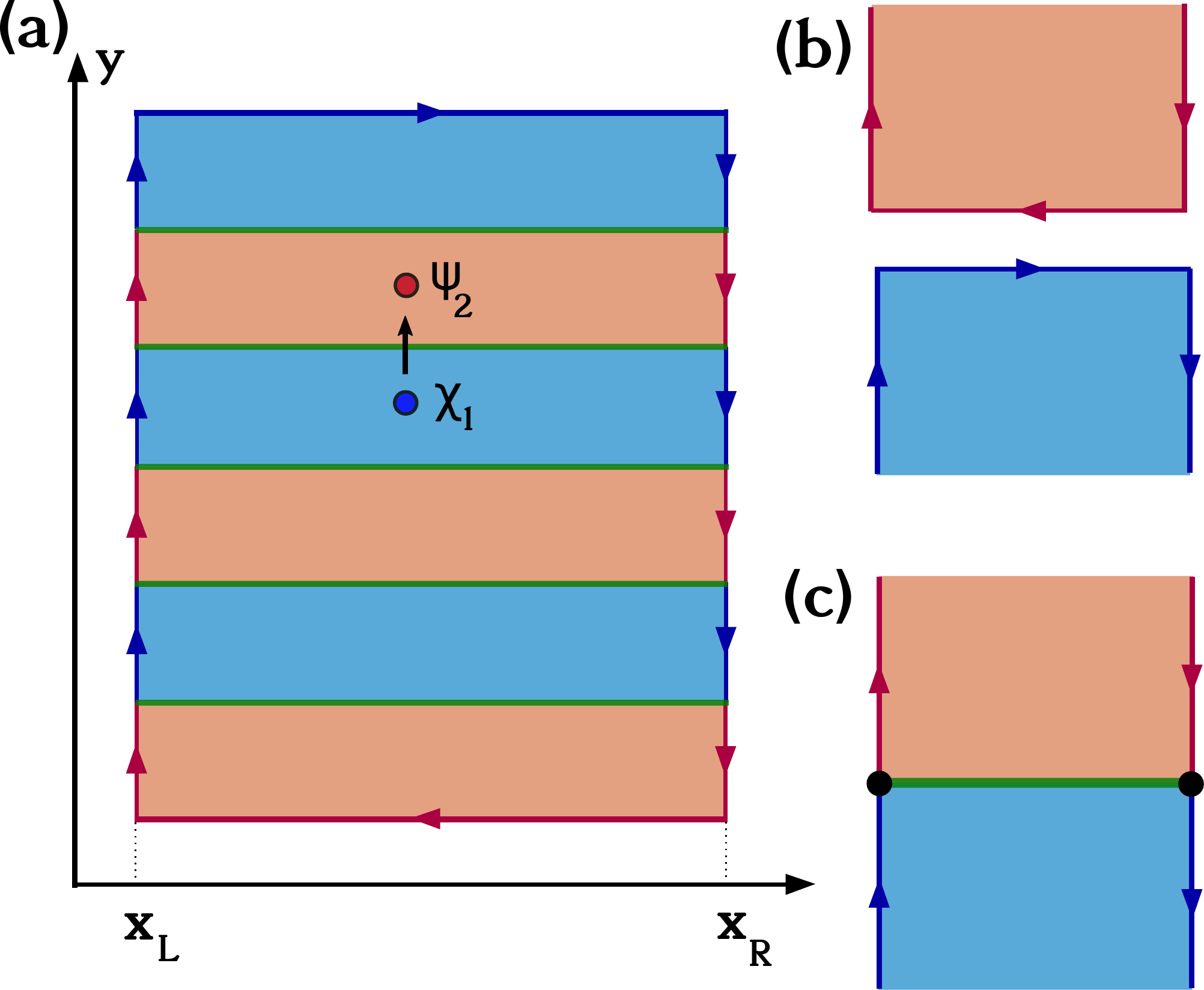}
\caption
{
(color online)
(a) An array of phases 1 (blue) and 2 (brown) showing the gapped interfaces
along the x direction (green). 
Each gapped interface constitutes an AC that acts as an anyon Andreev reflector
whereby certain quasiparticles of phase 1 are transformed into quasiparticles
of phase 2 (and vice-versa) as they cross the interface; and quasi-particles can be reflected into different quasiparticle types.
(b) Original chiral gapless edge states of the two phases.
(c) Parafermion zero modes (black dots) are located at the T-junctions
where the end points of the gapped interface define a domain wall
between the chiral gapless edge modes of phases 1 and 2.
}
\label{topological_interface}
\end{figure}

In this article we focus on the complementary effect that allows disparate 
TPs
to 
support \emph{gapped} interfaces (GIs), as they
provide a domain for a wide-range of interesting physics. 
The existence of such an interface requires that a local gapping condition be satisfied 
[see discussion around Eq.~\eqref{eq: null vector criterion}],
which, physically amounts to the allowed formation of an ``anyon condensate" (AC) at the interface.
It has been established, for two dimensional Abelian 
TPs,
that each AC is in one-to-one correspondence with a mathematical structure called a ``Lagrangian subgroup",~\cite{kapustin-2011,Levin-2013,Barkeshli-2013-a}
which is a subset $\mathcal{M}$ of the set of anyons wherein (i) all quasiparticles 
have mutual bosonic statistics,
and (ii) every quasiparticle not in $\mathcal{M}$ has non-trivial statistics with at least one quasiparticle of 
$\mathcal{M}$. Hence, the simultaneous condensation of the quasiparticles in $\mathcal{M}$ is allowed by (i), and will confine \emph{all} the anyons of the theory by (ii).
Of great interest are configurations where inequivalent ACs, corresponding to inequivalent choices of $\mathcal{M},$ are
formed in adjoining regions of a topological interface. Indeed, domain walls between these
gapped regions have been shown to host non-Abelian defect bound states with parafermionic 
statistics.\cite{Lindner-2012,Clarke-2013,Cheng-2012,Vaezi-2013,Barkeshli-2013-b,Mong-2014,Khan-2014,Khan-2016} 
Such bound states could be used as a platform for realizing topological quantum computation.~\cite{Nayak-2008}

In this letter we characterize a family of 1D gapped topological systems that can be formed at the 
interface between different 2D Abelian 
TPs.
For our examples, we choose single-component chiral phases characterized
by the topological invariants (K-matrices) $k_{1}$ and $k_{2},$ respectively.
Hereafter we refer to these as phase 1 and phase 2.
If these phases arise from charge conserving quantum Hall states then we have $k_{1,2} = \nu_{1,2}^{-1}$, where $\nu$ is the filling fraction that measures the Hall conductance in fundamental units.
More generally, for  systems
without U(1) (electromagnetic) charge conservation symmetry, e.g., 
chiral spin liquids~\cite{Kalmeyer-Laughlin-1987,Wen-Book},
$k_{1,2}$ count the number of distinct bulk quasiparticle
types in each phase, and give the topological ground state degeneracy 
$g^{k_{i}}$ of each system defined on a spatial manifold of genus $g$. 
For our discussion we will adopt the interface geometry in Fig. \ref{topological_interface}. The bulk 
TPs 
share a 
GI
with each other, and they have a boundary with the vacuum that contains propagating chiral edge modes,
such that the GI terminates at points separating gapless edge states of distinct phases.

Our main finding is that such an interface forms a topologically non-trivial, 1D gapped system with a
degenerate ground state manifold associated with parafermionic end states. %
We stress that, instead of being located at the 
domain walls between different GIs, 
the parafermions discussed here are situated at domain walls between
gapless edge states of phases 1 and 2, as shown in Fig.~\ref{topological_interface}.
Therefore this physical scenario departs significantly 
from those of Refs.~\cite{Lindner-2012,Clarke-2013,Cheng-2012,Vaezi-2013,Barkeshli-2013-b,Khan-2014,Khan-2016}, and more closely matches the setup of Ref. \onlinecite{Cano-2015}, though here we are focused more on what is happening in the bulk, rather than the edge as in their discussion. Ultimately, our results identify that, while one can find gapped interpolations between 2D phases with different topological order, these are not trivial gapped regions; they are instead \emph{topological} themselves.

We shall support our result with a bosonization
description of the edge containing a pair of counter propagating modes
from the two phases.
We will 
(1) construct the explicit form of the local, gap-opening interaction,  
(2) provide a description of the interface AC,  
(3) discuss the onset of the topologically degenerate ground state manifold
associated with the expectation value of a non-local operator,
and (4) discuss the connection between bulk confinement-deconfinement transitions, edge-state transitions, 
and the bound parafermion modes.


\textit{\textbf{1. Luttinger liquid description of the interface}} -- 
In Fig.~\ref{topological_interface}(a),
we consider an array of 2D topological states in phase 1 (blue) and phase 2 (brown), surrounded by the vacuum. As shown in the Supplementary Material (SM), the most generic gappable interface for one-component states is characterized
by  $k_{1} = p n^{2}$ and $k_{2} = pm^2,$ where $p, m, n \in \mathbb{Z}_{+}.$ 
The low energy Lagrangian of each interface along the x-direction
is given by
\begin{subequations}
\begin{equation}
\label{eq: interface Lagrangian}
\mathcal{L}_{x}
=
\frac{1}{4 \pi}\,
\partial_{t}\,\Phi^{T}\,K\,
\partial_{x}\,\Phi
-
\frac{1}{4 \pi}\,
\partial_{x}\,\Phi^{T}\,V\,
\partial_{x}\,\Phi
-
\mathcal{H}_{\textrm{int}}[\Phi]
\,,
\end{equation}
\begin{equation}
K
=
\begin{pmatrix}
p\,n^{2} & 0
\\
0 & -pm^2
\end{pmatrix}
\,,
\quad
\Phi
=
\begin{pmatrix}
\phi_{1}
\\
\phi_{2}
\end{pmatrix}
\,,
\end{equation}
\end{subequations}
where $\phi_{1,2}$ represent the right- and left-moving edge modes originating from phases 1 and 2,
V is a velocity matrix, and $\mathcal{H}_{int}[\Phi]$ is a local interaction 
discussed below.
The edge fields satisfy commutation relations
$[ \partial_{x}\,\phi_{i}(t,x), \phi_{j}(t, y) ] = - 2\,\pi\,\mathrm{i}\,K^{-1}_{i j}\,\delta(x - y).$
To simplify our discussion we will choose $m=1$ and provide the details for $m>1$ in the SM. This case is also the most experimentally relevant as it includes interfaces between a $\nu=1$ integer quantum Hall state with, e.g., a $\nu=1/9$ fractional quantum Hall state when $p=1, n=3.$ \cite{Pan-2002,Pan-2008}

With appropriate conventions,
the quasiparticle excitations on the edge are created by the  vertex operators
$\exp{\left(\mathrm{i}\,\ell^{T}\,\Phi\right)}$, where $\ell$ is an 
integer
vector.
The exchange statistics associated with taking a quasiparticle 
$\ell_a$ adiabatically around another quasiparticle $\ell_b$ is given by the statistical phase 
$S_{a b} = e^{\mathrm{i}\,\theta_{a b}} = e^{\mathrm{i}\,2\pi\,\ell^{T}_{a}\,K^{-1}\,\ell_{b}},$
and the (topological) spin of each quasiparticle is given by the self-statistics phase
$h_{a} = e^{\mathrm{i}\,\pi\,\ell^{T}_{a}\,K^{-1}\,\ell_{a}}$. 
Local excitations 
are identified with 
$\psi = e^{\mathrm{i}\,\Lambda^{T}\,K\,\Phi}$,
where $\Lambda$ is an integer vector.

In Eq. \eqref{eq: interface Lagrangian} 
$
\mathcal{H}_{\textrm{int}}[\Phi]
=
-J\,\cos{\left( \Lambda^{T}\,K\,\Phi \right)}
$
$
(J > 0)
$
is a local gap opening interaction 
parametrized by the integer null vector $\Lambda^{T} = (a, b)$ satisfying:~\cite{Haldane-1995}
\begin{equation}
\label{eq: null vector criterion}
0 = \Lambda^{T}\,K\,\Lambda =  p\,(a^{2}\,n^{2} - b^{2})
\,.
\end{equation}
$\Lambda = (1,n)$ is a \textit{primitive} solution\cite{LevinStern-2012} of \eqref{eq: null vector criterion}
representing the interaction between a single local operator $\psi_{1} = e^{\mathrm{i}\,p\,n^{2}\,\phi_{1}}$
of phase 1 with $n$ local operators $\psi^{\dagger}_{2} = e^{-\mathrm{i}\,p\,\phi_{2}}$ of phase 2:
\begin{equation}
\label{eq: strongly coupled Hamiltonian}
\begin{split}
\mathcal{H}_{\textrm{int}}
=
-J \cos{ \left( n \Theta \right) }
\propto
-J\,\psi_{1}\,
\underbrace
{
\psi^{\dagger}_{2}\,...\,\psi^{\dagger}_{2}
}
_
{
n	
}
+
\textrm{h.c.}
\,,
\end{split}
\end{equation}
where $\Theta(x) \equiv p\,n\,\phi_{1}(x) - p\,\phi_{2}(x)$. We explicitly show in the SM that one can always tune $V$ to make interactions of the form $\mathcal{H}_{int}$ relevant.

This interaction generates an AC at the interface as we will now describe.  
In phase 1 (phase 2), there are $p\,n^{2}$ ($p$) quasiparticle-types labeled $\varepsilon_{1}^{a_{1}}$
($\varepsilon_{2}^{a_{2}}$), $a_{1} = 1, ..., p\,n^{2}$ ($a_{2} = 1, ..., p$).
The set of anyons forms a discrete lattice\cite{Read-1990,Frohlich-1991,Wen-1992,Frohlich-1994,Cano-2014,Khan-2014},
whereby anyons are topologically indistinguishable upon the attachment
of local quasiparticles $\psi_{i} = \varepsilon_{i}^{k_{i}}$, $i = 1,2.$
In the context of a Laughlin fermionic (bosonic) state, $\psi_{i}$ represents the local 
fermion (boson) of the $i$-th phase.

Now we note that in phase 1 the anyon subset $\{ \varepsilon_{1}^{p\,n\,x}, x = 1, ..., n \}$
contains mutual bosons or fermions with spin $h(\varepsilon_{1}^{p\,n\,x}) = e^{\mathrm{i} \pi p x^{2}}$.
Furthermore,
the quasiparticle $\chi_{1} \equiv \varepsilon_{1}^{p\,n}$ 
has the same spin as the local excitation $\psi_{2}$ of phase 2, i.e., they are both bosons or fermions
depending on the parity of $p$. 
Physically this implies that the composite quasiparticle $\sigma\equiv \chi_{1}\,\psi^{\dagger}_{2}$ is a boson
that can condense, and generate a fully 
GI
between phases 1 and 2.
This condensation process, mathematically, is a consequence of 
the relation $k_{1}/k_{2} = n^{2} \in \mathbb{Z}^{2}$, which allows for
the existence of a $pn$-dimensional Lagrangian subgroup $\mathcal{M}$
containing $\sigma.$

Importantly, the interaction \eqref{eq: strongly coupled Hamiltonian}, which involves 
one local operator of phase 1 and $n$ of phase 2,
breaks the $U(1)\times U(1)$ particle conservation symmetries of each phase down to 
$\mathbb{Z}_1\times\mathbb{Z}_n$, where $\mathbb{Z}_1$ means no symmetry. 
Hence 
\eqref{eq: strongly coupled Hamiltonian} is invariant
under $S_{\beta}$: $\psi_{1} \rightarrow \psi_{1} $, 
$\psi_{2} \rightarrow \psi_{2} e^{\mathrm{i} 2\pi \beta/n}$, $\beta \in \mathbb{Z}.$ If the phases began with a $U(1)_{EM}$ electromagnetic charge conservation symmetry, then this interaction breaks (preserves) the symmetry when the charge vector is $t^{T} = (1,1)$ ($t^{T}=(n,1)$).
This discrete symmetry, it turns out, plays a fundamental role in the identification of the 
GI
as a topological parafermion wire similar to those studied in Refs. \onlinecite{Kitaev-2001,Fendley-2012,Bondesan-2013,Motruk-2013,Jermyn-2014,Zhuang-2015,Alexandradinata-2015}.

The topological properties of the 
GI
can be more transparently revealed by a description in the zero correlation length limit 
$J \rightarrow \infty$, 
where the interface Hamiltonian \textit{density} is given solely by Eq.~\eqref{eq: strongly coupled Hamiltonian},
thus leading to a 
GI
as depicted in Fig.~\ref{topological_interface}(c).
In this limit 
there are $n$ degenerate ground states
[$\Theta_{q} = 2\pi q / n$, $q = 1, ... , n$]
associated with the vacuum expectation value of the composite bosonic operator 
$
\sigma(x)
=
\chi_{1}(x)\,\psi^{\dagger}_{2}(x)
=
e^{\mathrm{i} \Theta(x)} 
$,
which represents a bound state of 
$
\chi_{1} 
= 
e^{\mathrm{i}\,p n \phi_{1}}
$ 
with 
$
\psi^{\dagger}_{2}
=
e^{- \mathrm{i}\,p \phi_{2}}
$:
\begin{equation}
\label{eq: eigenstates of sigma}
\forall x:
~~
\sigma(x)\,\ket{\Psi_{q}}
=
\omega^{q}\,\ket{\Psi_{q}}
\,,
~~
\omega \equiv e^{\mathrm{i}\frac{2\pi}{n}}
\,,
\quad
 q = 1, ..., n
\,.
\end{equation}

The eigenstates \eqref{eq: eigenstates of sigma} are in direct correspondence with symmetry broken ground
states of the ferromagnetic, zero correlation length limit of an
$n$-state clock model, where $\sigma$ naturally acquires the interpretation of a clock 
operator satisfying $\sigma^{n} = 1$ and $\sigma^{\dagger} = \sigma^{n-1}$.
However, while it would seem possible to distinguish among the degenerate states by a measurement
of $\sigma(x)$, $\bra{\Psi_{q}} \sigma(x) \ket{\Psi_{q'}} = \omega^{q} \delta_{q,q'}$
[which is equivalent to adding a perturbation $\delta\,\mathcal{H} = \delta\,\cos{(\Theta)}$
to the Hamiltonian \eqref{eq: strongly coupled Hamiltonian}],
the fact that $\sigma(x)$ is a \textit{non-local} operator does not permit such a local distinction,
and is a hallmark of the topological nature of the system.
With this in mind, the eigenstates \eqref{eq: eigenstates of sigma} indicate a 
degenerate symmetry breaking manifold associated with the global symmetry 
$
\mathcal{S}
\equiv
S_{(\beta = - 1)}
=
e^
{
-\frac{\mathrm{i}}{n}\,\int^{x_{R}}_{x_{L}} dx \, \partial_{x} \phi_{2}(x)		
}
$
whereby 
$
\mathcal{S}^{\dagger} \sigma(x) \mathcal{S} = \omega\,\sigma(x)
$,
for $x_{L} \leq x \leq x_{R}$.

The topological nature of this system can be made explicit by
changing from the clock to the parafermionic representation:~\cite{Fradkin-1980}
\begin{subequations}
\begin{equation}
\label{eq: def parafermion}
\alpha(x)
\equiv
\sigma(x)
\,
e^
{
-\frac{\mathrm{i}}{n}\,\int^{x}_{x_{L}} dz \, \partial_{z} \phi_{2}(z)		
}
\equiv
\sigma(x)
\,
\xi(x)
\,,
\end{equation}
\begin{equation}
\label{eq: parafermion algebra continuum}
\alpha(x)\,\alpha(y)
=
\alpha(y)\,\alpha(x)
\,
e^{\mathrm{i}\frac{2 \pi}{n} \textrm{sgn}(y - x)}
\,,
\end{equation}
\end{subequations}
whereby 
$\alpha(x)$ is a product 
of the ``order", $\sigma$, and the ``disorder", $\xi$, operators. 
Importantly, 
the boundary parafermion operators
$
\alpha(x_{L})
=
\sigma(x_{L})
\,,
~~
\alpha(x_{R})
=
\sigma(x_{R})
\,
e^
{
-\frac{\mathrm{i}}{n}\,\int^{x_{R}}_{x_{L}} dx' \, \partial_{x'} \phi_{2}(x')		
}
$
commute with the Hamiltonian \eqref{eq: strongly coupled Hamiltonian},
and the degenerate ground state manifold is given by the eigenstates of
the non-local operator 
$
\mathcal{A}
=
\alpha^{\dagger}(x_{L})\,\alpha(x_{R})
$:
$\mathcal{A} \ket{\Omega_{a}} = \omega^{a} \ket{\Omega_{a}}$,
$a = 1,\ldots,n,$ where the $\ket{\Omega_{a}}$ are linear combinations of the $\ket{\Psi_{q}}.$


\textit{\textbf{2. Edge transitions}} -- 
As indicated in Figs.~\ref{topological_interface}(b) and~\ref{topological_interface}(c),
the formation of the 
GI
prevents the propagation of the edge modes in the x-direction.
While any point $x \in (x_{L},x_{R})$ establishes a domain wall between distinct gapped bulk 
TPs,
the end states located at $x = x_{L,R}$ correspond to domain walls between distinct \emph{gapless} edge states.
In fact we shall explicitly demonstrate the existence of parafermion
operators situated at the edge transitions. These parafermions 
are non-trivial operators with quantum dimensions $\sqrt{n}$,
which is a direct manifestation of the $n$-fold degeneracy of the GI.
Similar physics was first explored in Ref.~\cite{Cano-2015}, which focuses on transitions between distinct edge terminations of 
the \emph{same} bulk phase; our focus instead is on the interface between \textit{different} bulk phases, which will have an accompanying transition on the edge.

An important feature of the gappable topological interface is that the bulk phases 1 and 2 can be related to each other
by the confinement (or deconfinement) of a 2D $\mathbb{Z}_{n}$ gauge theory.  	
In order to see this, imagine  phase 2  is coupled
to a $\mathbb{Z}_{n}$ gauge theory in its deconfined phase.
Let the gauge field $\alpha_{\mu}$
describe the excitations of phase 2, and ($a_{\mu},b_{\mu}$)
the excitations of the $\mathbb{Z}_{n}$ gauge theory. Hence,
the coupled system is described by the Abelian Chern-Simons
theory:
\begin{subequations}
\label{eq: 2D state coupled to Zn gauge theory}
\begin{equation}
\begin{split}
\mathcal{L}_{2D}
=
\frac{1}{4\pi}\,\varepsilon^{\mu\nu\lambda}\,
c^{I}_{\mu}
\,
\bar{K}_{I J}(p,n)
\,
\partial_{\nu}\,c^{J}_{\lambda}
\,,
\end{split}•
\end{equation}•
\begin{equation}
\label{eq: k matrix coupled to to Zn gauge theory}
\bar{K}(p,n)
=
\begin{pmatrix}
p & -1 & 0
\\
-1 & 0  & n
\\
0 & n & 0
\end{pmatrix}•
\,\quad
c_{\mu}
=
\begin{pmatrix}
\alpha_{\mu}
\\
a_{\mu}
\\
b_{\mu}
\end{pmatrix}•
\,,
\end{equation}•
\end{subequations}\noindent where $\mu,\nu,\lambda \in \{0,1,2\}$. In this basis $e = (0,1,0)$ and $m = (0,0,1)$ represent the original 
charge and flux excitations of the gauge theory.

A $W \in \textrm{GL}(3,\mathbb{Z})$ change of basis yields~\cite{Supplementary-Material}
\begin{equation}
\label{eq: rotated K-matrix}
K_{g}
\equiv
W^{T}\,\bar{K}(p,n)\,W
=
p\,n^{2} \oplus \Sigma
\,,
\end{equation}
where $\Sigma$ represents a Pauli matrix, i.e., a trivial sector that can always be gapped out. 
Thus, Eq.~\eqref{eq: rotated K-matrix} explicitly illustrates that phase 1 can be obtained from phase 2 by a gauging mechanism;
reversely, phase 2 descends from phase 1 by confining the $\mathbb{Z}_{n}$ gauge theory. 
This kind of gauging mechanism has proven useful in 
understanding the classification of symmetry enriched topological states~\cite{Lu-2016} and hidden anyonic symmetries\cite{Khan-2016}.

We now explicitly prove the existence of domain-wall parafermions
by analyzing the transitions between edge phases 1 and 2.
The transitions can be analyzed starting from the bulk theory in Eq. \eqref{eq: 2D state coupled to Zn gauge theory}, 
and using the standard bulk-boundary correspondence for Abelian topological phases~\cite{Wen-1990}. 
Hence, we model gapless edge states propagating along one of the edges, say $x = x_{L}$, with the effective theory	
\begin{equation}
\label{eq: y edge Lagrangian}
\begin{split}
\mathcal{L}_{x_{L},\textrm{y}}
&\,=
\frac{1}{4 \pi}
\partial_{t}\Phi^{'T}\,K_{g}
\partial_{y}\Phi'
+
\sum^{2}_{a=1}J_{a}(y)\,H_{int,a}
\end{split}
\end{equation}
where 
$\Phi^{'T}(t,x_{L},y) = (\phi^{'}_{1},\phi^{'}_{2},\phi^{'}_{3})(t,x_{L},y)$
are the edge fields. 
The interactions $H_{int,1}$ and $H_{int,2}$ will be chosen to stabilize the edge phases 1 and 2, respectively, in different spatial regions, i.e., the interaction $H_{int,1}$ ($H_{int,2}$) partially gaps out two of the three edge modes to leave the single-component edge mode of phase 1 (phase 2).
To carry this out we use position-dependent coupling constants $J_{1}(y)$ and $J_{2}(y)$ such
that, $J_{1} \rightarrow \infty$ and $J_{2} = 0$ in phase 1, while $J_{1}=0$ and $J_{2} \rightarrow \infty$ in phase 2.
For concreteness, we take $p = 2\,q+1$ and $\Sigma = \sigma_{z}$
in \eqref{eq: rotated K-matrix}, although similar results can be obtained for the $p = 2\,q$ case with $\Sigma=\sigma_{x}.$

The interaction choice
\begin{equation}
\label{eq: interaction V1}
H_{int,1}
=
\cos
{
\left(
L_{1}^{T}\,K_{g}\,\Phi'
\right)
}
\,,
\quad
L^{T}_{1} = (0,1,1)
\,
\end{equation}
will gap the trivial modes in $\Sigma$ yielding the edge states of phase $1$.
Alternatively, the interaction
\begin{equation}
\label{eq: interaction V2}
H_{int,2}
=
\cos
{
\left(
L_{2}^{T}\,K_{g}\,\Phi'
\right)
}
\,,
\quad
L^{T}_{2} = (1,q\,n,(q+1)\,n)
\,
\end{equation}
gives rise to the edge state of phase 2, that is,
it effectively leads to the confinement of the $\mathbb{Z}_{n}$ gauge theory.
To see this, notice that the edge excitations that remain deconfined
in the presence of the interaction \eqref{eq: interaction V2}
are described by vertex operators
$\exp{(\mathrm{i}\,\ell^{T}\,\Phi^{'})}$, 
with $\ell^{T} = (\ell_{1},\ell_{2},\ell_{3})$,
such that $\ell^{T}\Phi^{'}$
commutes with the argument of the interaction \eqref{eq: interaction V2}.
From this condition, which is satisfied when $\ell_{1} = -n\,\left[ \ell_{2}\,q  + (q+1)\ell_{3} \right]$,
we find that the deconfined edge excitations are those of the phase 2 described by $k_{2} = p = (2\,q+1)$.
More intuitively, upon rewriting
$
H_{int,2}
=
\cos
{
\left(
L_{2}^{T}\,K_{g}\,\Phi'
\right)
}
=
\cos
{
\left(
\bar{L}_{2}^{T}\,\bar{K}\,\bar{\Phi}'
\right)
}
=
\cos
{
\left(
n\,\bar{\phi}^{'}_{2}
\right)
}
$,
with $\bar{L} = W L$ and $\bar{\Phi}^{'} = W \Phi^{'}$,
\eqref{eq: interaction V2} is seen as the expected  ``electric"-mass
interaction that confines the excitations of the $\mathbb{Z}_{n}$ gauge theory.

Defining the segments $R^{\pm}_{1,i} = \left( y_{2i-1} \pm \varepsilon, y_{2i} \mp \varepsilon \right)$
and $R^{\pm}_{2,i} = \cup_{i} \left( y_{2i} \pm \varepsilon, y_{2i+1} \mp \varepsilon \right)$,
$\varepsilon = 0^{+}$, we let the regions $R^{+}_{a} = \cup_{i}\,R^{+}_{a,i}$, with $a=1,2$,
denote the edge phases 1 and 2 along the $x = x_{L}$ edge.
The operators 
$
\mathcal{O}^{(a)}_{i}
=
\exp
\Big[
{
\frac{\mathrm{i}}{n}\,
\int_{R^{-}_{a,i}}\,dy\,\partial_{y} \left( L^{T}_{\bar{a}}\,K_{g}\,\Phi'\right)
\Big]
},
$
where $\bar{a} \equiv a + (-1)^{a+1}$,
are seen to commute with the edge Hamiltonian and satisfy the non-trivial commutation relations
\begin{equation}
\label{eq: algebra of O}
\begin{split}
\mathcal{O}^{(1)}_{i}\,\mathcal{O}^{(2)}_{k} = \mathcal{O}^{(2)}_{k}\,\mathcal{O}^{(1)}_{i}\,
e^{\frac{2\pi \mathrm{i}}{n}\left(\delta_{k,i-1} - \delta_{k,i} \right)}
\,.
\end{split}
\end{equation}
The ground state manifold forms a representation of the algebra \eqref{eq: algebra of O},
which implies a ground state degeneracy of $n^{k-1}$ in the presence of $2\,k$ domain walls on the boundary, i.e.,
$k$ GIs.
The operators
\begin{equation}
\label{eq: parafermions definition and algebra}
\begin{split}
&\,
\alpha_{x_{L},\,{\ell}}
=
e^
{
\frac{\mathrm{i}}{n}\,
\Big[
L^{T}_{a_{\ell}} K_{g} \Phi^{'}(y_{\ell} + \varepsilon)
-
L^{T}_{\bar{a}_{\ell}} K_{g} \Phi^{'}(y_{\ell} - \varepsilon)
\Big]		
}
\,,
\end{split}
\end{equation}
[$a_{2i\,(2i+1)} \equiv 2\,(1)$]
with support on the domain walls along the $x = x_{L}$ edge satisfy, as expected,
parafermionic algebra
$
\alpha_{x_{L},k}\,\alpha_{x_{L},\ell} 
= 
\alpha_{x_{L},\ell}\,\alpha_{x_{L},k}\,
\omega^{\textrm{sgn}(k - \ell)\,(-1)^{k+\ell}}.
$
For a generic GI between one-component states we have the constraint $k_1=\tfrac{n^2}{m^2}k_2$ which implies that the phases must be related by the confinement of a $\mathbb{Z}_m$ gauge theory, and the subsequent gauging and deconfinement of a $\mathbb{Z}_n$ 
symmetry. In these cases one would find $\mathbb{Z}_{mn}$ parafermions (see SM for more detail).

A realization of the algebra \eqref{eq: algebra of O}
has been studied in Ref.~\cite{Cano-2015}, for the transitions
between chiral bosonic edge states with $k_{1} = 2\,n^{2}$ and $k_{2} = 2$.
While their approach focused solely on the edge transitions of a homogenous bulk phase, our formulation shows that the existence of non-trivial parafermionic modes \eqref{eq: parafermions definition and algebra} is a direct consequence of the formation of a 
GI
between different chiral topological states. 
Hence, we have generalized their result to arbitrary one-component edge transitions, and have shown that such transitions can originate from a bulk phenomenon associated with confinement-deconfinement transitions of discrete gauge theories.
Additionally, since these parafermions appear
at a ``T-junction" between two chiral gapless states
and the termination of their 
GI,
they represent a completely new physical phenomenon when compared with the cases studied in 
Refs.~\cite{Lindner-2012,Clarke-2013,Cheng-2012,Vaezi-2013,Barkeshli-2013-b,Khan-2014,Khan-2016}. 

We note that the 
GI
acts like an anyonic Andreev reflector in the bulk. 
Anyons from, say, phase 1 will hit the interface and be transformed into a combination of outgoing anyons in phase 2 as well as reflected anyons that remain in phase 1. Take $p = 1, m=1$ for simplicity.
Then as, for example, quasiparticle $\chi_{1} = \varepsilon^{n}_{1}$ approaches the interface,
a vacuum fluctuation can create a ($\psi_{2},\bar{\psi}_{2}$)
pair in the region of phase 2 immediately adjacent to the interface;
subsequently, the condensation of ($\chi_{1} \bar{\psi}_{2}$)
leaves behind the quasiparticle $\psi_{2}$ in phase 2,
as shown in Fig.~\ref{topological_interface}(a).
The quasiparticles $\{ \varepsilon_{1}^{n\,x}, x \in \mathbb{Z} \}$ belonging
to phase 1 can be absorbed by the 
GI
and fully transmuted into multiples
of the local excitation $\psi_{2}$ of phase 2. Other anyons hitting the interface will be partially transmuted and partially reflected by the condensate.
For example, if $\varepsilon_{1}$ hits the surface it could generate a 
$\psi_{2}$ in phase 2 as well as a reflected $\varepsilon^{(-n+1)}_{1}.$

In summary, we have shown that a gapped interface between different topologically ordered phases cannot be topologically trivial itself. The interpolation between the topological orders generates a quasi-1D topological parafermion phase which exhibits characteristic non-Abelian defect modes where the interface intersects the boundary of the system. Although we have only shown this for one-component interfaces, we expect the generalizations to more complicated interfaces to provide a rich set of phenomena. Furthermore, our result may aid in the interpretation of the topological entanglement entropy arising at heterointerfaces of topologically ordered phases as recently calculated in Ref. \onlinecite{canohughesmulligan}. We leave this to future work.

\begin{acknowledgements}
We would like to thank J. Cano, E. Fradkin, M. Mulligan and M. Stone for useful conversations.
LHS is supported by a fellowship from the
Gordon and Betty Moore Foundation's EPiQS Initiative through
Grant No. GBMF4305 at the University of Illinois.
TLH is supported by the US National Science Foundation under grant DMR 1351895-CAR.
\end{acknowledgements}

\pagebreak
\widetext
\begin{center}
\textbf{\large Supplementary Material of \textit{Parafermionic wires at the interface of chiral topological states}}
\end{center}
\setcounter{equation}{0}
\setcounter{figure}{0}
\setcounter{table}{0}
\setcounter{page}{1}
\makeatletter
\renewcommand{\theequation}{S\arabic{equation}}
\renewcommand{\thefigure}{S\arabic{figure}}
\renewcommand{\bibnumfmt}[1]{[S#1]}
\renewcommand{\citenumfont}[1]{S#1}


\section{Gapability Condition and Parafermions}
\label{sec: Gapability Condition and Parafermions}

Suppose we have an interface between two chiral, one-component theories with K-matrices\cite{Wen-1990-suppl} 
$k_1$ and $k_2$ respectively, which are both positive integers. The K-matrix at the interface will be treated as $K={\rm{diag}}[k_1, -k_2].$ The effective edge theory reads 
\begin{subequations}
\begin{equation}
\label{eq: interface Lagrangian k1 k2 - supplement}
\mathcal{L}_{x}
=
\frac{1}{4 \pi}\,
\Big(
\partial_{t}\,\Phi^{T}\,K\,\partial_{x}\,\Phi
-
\partial_{x}\,\Phi^{T}\,V\,\partial_{x}\,\Phi
\Big)
-
\mathcal{H}_{\textrm{int}}[\Phi]
\,,
\end{equation}
\begin{equation}
K
=
\begin{pmatrix}
k_1 & 0
\\
0 & -k_2
\end{pmatrix}
\,,
\quad
\Phi
=
\begin{pmatrix}
\phi_{1}
\\
\phi_{2}
\end{pmatrix}
\,,
\quad
\mathcal{H}_{\textrm{int}}[\Phi] = -J\,\cos{\left( \Lambda^{T}\,K\,\Phi \right)}
\,,
\end{equation}
\end{subequations}
where $\phi_{1,2}$ represent the excitations originating from the edges states of phases 1 and 2, with equal-time commutation relations
$
[ \partial_{x}\,\phi_{i}(t,x), \phi_{j}(t, y) ] = - 2\,\pi\,\mathrm{i}\,K^{-1}_{i j}\,\delta(x - y)
$, and $\mathcal{H}_{\textrm{int}}[\Phi]$ is a gap opening interaction.

The null vector criterion,\cite{Haldane-1995-suppl} for a null vector $\Lambda= (a,b)$ with $a,b\in\mathbb{Z}$ is
\begin{equation}
\label{eq:gapping}
a^2 k_1=b^2 k_2.
\end{equation} 
Let us now consider this equation in its generality. To, simplify let $p={\rm{gcd}}(k_1, k_2)$ such that $k_1=p \bar{k}_1$ and 
$k_2=p\bar{k}_2.$ We will provide a simple physical interpretation for $p$ later, but for now it serves to simplify the gapability constraint. We can reformulate Eq. \ref{eq:gapping} as\begin{equation}
\frac{k_1}{k_2}=\frac{\bar{k}_1}{\bar{k}_2}=\left(\frac{b}{a}\right)^2.\label{eq:gapping2}
\end{equation} We immediately see that $p$ plays no role in the determination of this condition, and by construction $\bar{k}_1$ and $\bar{k}_2$ are relatively prime. 

We can now multiply through to find
\begin{equation}
a^2\bar{k}_1=b^2\bar{k}_2
\end{equation} 
and this implies, using the relatively prime condition, that $\bar{k}_1$ divides $b^2$ and that $\bar{k}_2$ divides $a^2.$ As such we have the restrictions
\begin{eqnarray}
a^2&=&\bar{k}_2\gamma_a\nonumber\\
b^2&=&\bar{k}_1\gamma_b
\end{eqnarray} where $\gamma_a$ and $\gamma_b$ are integers. Plugging this into Eq. \ref{eq:gapping2} we arrive at $\gamma_a=\gamma_b.$

To summarize, we now have the relations
\begin{eqnarray}
k_{1}&=& p\,\bar{k}_1,\;\;\; k_{2}= p\,\bar{k}_2\nonumber\\
a^2&=& \bar{k}_2 \gamma_a,\;\;\; b^2= \bar{k}_1 \gamma_a.
\end{eqnarray} 
Now, in order for $\Lambda$ to be a primitive vector, $a$ and $b$ cannot have common factors, hence $\gamma_a=1$ if we enforce primitivity.   Hence, we arrive at the result that the most generic gappable interface for two, one-component theories is 
$K={\rm{diag}}[ p\,n^2, -p\,m^2].$ Furthermore, a primitive gapping vector that will gap this theory takes the form 
$\Lambda=(m,n).$

With this result we can gain a physical understanding. If we take the theory $k_1$ and gauge a 
$\mathbb{Z}_m$ subgroup of the global particle number conservation symmetry and take $k_2$ and gauge a 
$\mathbb{Z}_n$ subgroup of its particle number conservation symmetry then we will arrive at two theories with the same topological order, and which can hence be gapped. In fact, the gapability equation written as $k_1=\tfrac{b^2}{a^2}k_2$ essentially encodes that if we take the theory $k_2$, confine a $\mathbb{Z}_a$ group and deconfine a $\mathbb{Z}_b$ group then we will arrive at the same topological order as $k_1.$ From our constraints we see this is exactly true since $a^2$ divides $k_2.$ The common factor 
$p$ is interpreted as a common parent theory from which each topological phase $k_1, k_2$ can be reached by deconfining discrete 
$\mathbb{Z}_n$ and $\mathbb{Z}_m$ subgroups of $U(1)$ respectively.

Let us now look at the condensate at the interface. The gapping term can be written as
\begin{equation}
\label{eq: strongly coupled Hamiltonian general case k1 and k2- supplement}
\begin{split}
H_{int}
&\,=
-J\,
\int^{x_{R}}_{x_{L}} dx
\cos{ \Big[ m n \left( p n \phi_{1} -  p m \phi_{2} \right) \Big]}
\equiv
-J\,
\int^{x_{R}}_{x_{L}} dx
\cos{ \left( m n \Theta \right) }
\\
&\,
\propto
-J\,
\int\,dx\, \left( \psi_{1} \right)^{m}\,\left( \psi^{\dagger}_{2} \right)^{n} + \textrm{H.c.}
\,.
\end{split}
\end{equation}
It is important not to factor out $p$ when considering the condensate. To see this let us consider an example for the case when $k_1=2$ and $k_2=8$ where $p=2$, $n=1$ and $m=2.$ In phase $1$ the statistics are $\exp(i\pi r^2/2)$ where $r\in \mathbb{Z}.$ The only fermion or boson in the quasiparticle set is when $r$ is an even integer, i.e., the local boson in this case, which we will call $\psi_1.$ For phase $2$ the statistics are $\exp(i\pi r^2/8)$ and the quasiparticle with $r=4$ is a (non-local) boson which we call $\chi_2$. For this system we can form $(\psi_1\chi_2^{\dagger})^2=\psi_{1}^2\psi_{2}^{\dagger}$ to get a local condensate. 
The interaction that will generate this condensate is $J\cos\left(4\phi_1-8\phi_2 \right)=J\cos(2(2\phi_1-4\phi_2)).$ From the form of the condensate we see that the $U(1)\times U(1)$ group structure is broken to $\mathbb{Z}_{2}\times\mathbb{Z}_1,$ and hence we expect there to be $\mathbb{Z}_2$ parafermions. In the generic case the symmetry is broken to $\mathbb{Z}_{n}\times \mathbb{Z}_m$ and hence a $\mathbb{Z}_{mn}$ parafermion results. 
To see this explicitly, note that the second line of \eqref{eq: strongly coupled Hamiltonian general case k1 and k2- supplement}
makes explicit that the interaction pairs $m$ operators of phase 1 with $n$ operators of phase 2.
The non-local operator $\sigma(x) = e^{\mathrm{i} \Theta(x)}$, where 
$\Theta(x) \equiv p\,n\,\phi_{1}(x) - p\,m\,\phi_{2}(x)$ shall be identified with a $\mathbb{Z}_{m n}$ clock variable.

From the form of this interaction, we can immediately write down three symmetry operations
and their representations in terms of the bosonic fields:
\begin{equation}
\begin{split}
&\,
\tilde{S}_{\alpha}:
\quad
\psi_{1} \rightarrow \psi_{1}\,e^{\mathrm{i} 2\pi n \alpha}
\,,~
\psi_{2} \rightarrow \psi_{2}\,e^{\mathrm{i} 2\pi m \alpha}
\,,
\quad \alpha \in \mathbb{R}
\,, 
\\
&\,
\tilde{S}_{\alpha}
=
e^
{
-\mathrm{i}\,\alpha \int^{x_{R}}_{x_{L}} dx\,
\left(
n \partial_{x} \phi_{1}
-
m \partial_{x} \phi_{2}
\right)
}
\,,
\end{split}
\end{equation}
which is a U(1) symmetry of the interaction Hamiltonian when the topological phases
have charge vector $t^{T} = (n,m)$. 

Moreover, we have
\begin{equation}
\begin{split}
&\,
S_{1,\beta_{1}}:
\quad
\psi_{1} \rightarrow \psi_{1}\,e^{-\mathrm{i} 2\pi \beta_{1}/m}
\,,~
\psi_{2} \rightarrow \psi_{2}
\,,
\quad \beta_{1} \in \mathbb{Z}_{m}
\,,
\\
&\,
S_{1,\beta_{1}}
=
e^
{
\mathrm{i}\,(\beta_{1}/m) \int^{x_{R}}_{x_{L}} dx\,
\partial_{x} \phi_{1}
}
\,,
\end{split}
\end{equation}
which accounts for a discrete $\mathbb{Z}_{m}$ associated with phase 1 and, equivalently,
\begin{equation}
\begin{split}
&\,
S_{2,\beta_{2}}:
\quad
\psi_{1} \rightarrow \psi_{1}
\,,~
\psi_{2} \rightarrow \psi_{2}\,e^{\mathrm{i} 2\pi \beta_{2}/n}
\,,
\quad \beta_{2} \in \mathbb{Z}_{n}
\,,
\\
&\,
S_{2,\beta_{2}}
=
e^
{
\mathrm{i}\,(\beta_{2}/n) \int^{x_{R}}_{x_{L}} dx\,
\partial_{x} \phi_{2}
}
\,,
\end{split}
\end{equation}
which accounts for a discrete $\mathbb{Z}_{n}$ associated with phase 2.

Importantly, while $\sigma(x)$ does not transform under the action of $\tilde{S}_{\alpha}$, 
it does so under the other two operators as follows:
\begin{equation}
S^{\dagger}_{1,\beta_{1}} \, \sigma \, S_{1,\beta_{1}} = \sigma\,e^{-\mathrm{i} 2\pi \beta_{1}/mn}
\,,~
S^{\dagger}_{2,\beta_{2}} \, \sigma \, S_{2,\beta_{2}} = \sigma\,e^{-\mathrm{i} 2\pi \beta_{2}/mn}
\end{equation}

Therefore in the clock representation, the ground state degeneracy of the 
Hamiltonian \eqref{eq: strongly coupled Hamiltonian general case k1 and k2- supplement}
becomes associated with the ``symmetry broken" states
\begin{equation}
\label{eq: eigenstates of sigma - supplement}
\forall x:
~~
\sigma(x)\,\ket{\Psi_{q}}
=
\omega^{q}\,\ket{\Psi_{q}}
\,,
\quad
 q = 1, ..., m n
\,,
\end{equation}
where $\omega \equiv e^{\mathrm{i}\frac{2\pi}{m n}}$.
While it would seem that a particular eigenstate could be identified by the local measurement
of $\sigma(x)$, $\bra{\Psi_{q}} \sigma(x) \ket{\Psi_{q'}} = \omega^{q}\,\delta_{q,q'}$,
it is fundamental to recognize that $\sigma(x)$ is \textit{not} a local operator in terms of the 
original local degrees of freedom $\psi_{1}$ and $\psi_{2}$.

\subsection{Relevance of the gap opening interaction \eqref{eq: strongly coupled Hamiltonian general case k1 and k2- supplement}}

We consider the edge theory \eqref{eq: interface Lagrangian k1 k2 - supplement},
with $K = \textrm{diag}[p\,n^{2}, -p\,m^{2}]$ ($p$, $m$ and $n$ positive integers), 
$\Lambda = (m,n)$ the null vector appearing in \eqref{eq: strongly coupled Hamiltonian general case k1 and k2- supplement},
and $V$ the positive definite matrix that parametrizes the forward scattering of the edge modes. 
We now provide an example of a $V$ matrix for which the interaction $\mathcal{H}_{int} = -J\,\cos{\left( \Lambda^{T}\,K\,\Phi \right)}$ is a relevant operator.

First, perform a change of variables $\Phi = M\,\Phi'$, where $M = \textrm{diag}[(p\,n^{2})^{-1/2}, -(p\,m^{2})^{-1/2}]$,
which rescales $K' = M^{T}\,K\,M = \sigma_z$ and $V' = M^{T}\,V\,M$. Define another change of basis
$\Phi' = O\,\tilde{\Phi}$ that diagonalizes $\tilde{V} = O^{T}\,V'\,O = \textrm{diag}[\tilde{v}_1, \tilde{v}_2]$ 
($\tilde{v}_{1,2} > 0$) while keeping $K'= O^{T}\,K'\,O=\sigma_z$ unchanged. The interaction, in this new basis, reads
\begin{equation}
\mathcal{H}_{int} = -J\,\cos{\left( \tilde{a}^{T}\,\tilde{\Phi} \right)}
\,,
\textrm{with}~
\tilde{a} = O^{T}a
\,\,
\textrm{and}~
a =
\begin{pmatrix}
\sqrt{p}\,m\,n
\\
- \sqrt{p}\,m\,n
\end{pmatrix}
\,.
\end{equation}

From the correlator
\begin{equation}
\Big\langle
e^{\mathrm{i}\,\tilde{a}^{T}\,\tilde{\Phi}(t,x)}
\,
e^{-\mathrm{i}\,\tilde{a}^{T}\,\tilde{\Phi}(0,0)}	
\Big\rangle
=
\frac{1}{\left( \tilde{v}_{1}\,t - x \right)^{\tilde{a}^{2}_{1}}}
\,
\frac{1}{\left( \tilde{v}_{2}\,t + x \right)^{\tilde{a}^{2}_{2}}}
\,,
\end{equation}
the scaling dimension of the interaction can be extracted:
\begin{equation}
\Delta = \frac{1}{2}\,\tilde{a}^{T}\,\tilde{a} = \frac{1}{2}\,a^{T}\,O O^{T}\,a = \frac{1}{2}\,a^{T}\,B^{2}\,a
\,.
\end{equation}
The last equality was obtained after expressing $O = B\,R$, where $B$ and $R$ are, respectively,
a symmetric positive boost and an orthogonal rotation, $R\,R^{T} = I$.\cite{Moore-1998-suppl,Moore-2002-suppl}
An example of a transformation $O$ and a matrix $V'$ that satisfy these properties is given by\cite{Khan-2016-suppl}
\begin{subequations}
\begin{equation}
O = B = e^{\frac{b}{2}\,\sigma_x}
\,,
\quad
V' = e^{-b\,\sigma_x}
\,,
\quad b \in \mathbb{R}
\,,
\end{equation}
for which
\begin{equation}
O^{T}\,V'\,O = \tilde{V} = 
\begin{pmatrix}
1 & 0
\\
0 & 1
\end{pmatrix}	
\,,
\quad
O^{T}\,\sigma_{z}\,O = \sigma_{z}
\,.
\end{equation}
\end{subequations}

This parametrization yields the scaling dimension $\Delta = p\,m^{2}\,n^{2}\,e^{-b}$. The condition $\Delta < 2$, 
required for the gap opening interaction to be a relevant operator acting on the $(1+1)$ dimensional interface, is satisfied if
$b > \log{\Big( \frac{p\,m^{2}\,n^{2}}{2} \Big)}$.

\section{Gapped Interface for the $n=1$ case}
\label{sec: Gapped Interface for the m=1 case}

We show the existence of an operator that commutes with the interface Hamiltonian
\begin{equation}
\label{eq: strongly coupled Hamiltonian - supplement}
\begin{split}
&\,
H_{int}
=
\int^{x_{R}}_{x_{L}} dx\,
\mathcal{H}_{\textrm{int}}
=
-J\,
\int^{x_{R}}_{x_{L}} dx 
\cos{ \left( n \Theta \right) }
\,,
\quad
\Theta(x) \equiv p\,n\,\phi_{1}(x) - p\,\phi_{2}(x)
\,
\end{split}
\end{equation}
and yields an $n$-fold degeneracy of the ground state of \eqref{eq: strongly coupled Hamiltonian - supplement}.
For that we seek a unitary transformation
\begin{equation}
\Sigma(a_{1},a_{2})
=
e^
{
\mathrm{i}\,
\left[
a_{1}\,\int^{x_{R}}_{x{L}}\,dx\,\partial_{x}\,\phi_{1}(x)
+
a_{2}\,\int^{x_{R}}_{x{L}}\,dx\,\partial_{x}\,\phi_{2}(x)
\,
\right]
}
\,
\end{equation}
parametrized by $a_{1,2} \in \mathbb{R}$ that commutes with the 
Hamiltonian \eqref{eq: strongly coupled Hamiltonian - supplement}.
With the equal time commutation relations 
$[ \partial_{x}\,\phi_{1}(x), \phi_{1}(y) ] = -\frac{2\,\pi\,\mathrm{i}}{p\,n^{2}}\,\delta(x - y)$,
$[ \partial_{x}\,\phi_{2}(x), \phi_{2}(y) ] = \frac{2\,\pi\,\mathrm{i}}{p}\,\delta(x - y)$ and
$[ \partial_{x}\,\phi_{1}(x), \phi_{2}(y) ] = [ \partial_{x}\,\phi_{2}(x), \phi_{1}(y)] = 0$,
we find that
$
\Sigma^{\dagger}(a_{1},a_{2})\,\phi_{1}(x)\,\Sigma^{}(a_{1},a_{2}) = \phi_{1}(x) - \frac{2 \pi a_{1}}{p n^{2}}
$
and
$
\Sigma^{\dagger}(a_{1},a_{2})\,\phi_{2}(x)\,\Sigma^{}(a_{1},a_{2}) = \phi_{2}(x) + \frac{2 \pi a_{2}}{p}
$,
where $x_{L} \leq x \leq x_{R}$.  With that the local operators 
$\psi_{1}(x) = e^{\mathrm{i} p n^{2} \phi_{1}(x)}$ and $\psi_{2}(x) = e^{\mathrm{i} p \phi_{2}(x)}$
transform as:
$
\Sigma^{\dagger}(a_{1},a_{2})
\,
\left(
\psi_{1}(x), \psi_{2}(x)
\right)
\,
\Sigma^{}(a_{1},a_{2})
=
\left(
\psi_{1}(x)\,e^{-\mathrm{i} 2 \pi a_{1}}, \psi_{2}(x)\, e^{+\mathrm{i} 2 \pi a_{2}}
\right)
$.
It then follows
\begin{equation}
\label{eq: condition on Sigma for invariance of H - supplement}
[\Sigma^{\dagger}(a_{1}, a_{2}), H_{int}]  = 0 
\quad
\Longleftrightarrow
\quad
a_{1} + n a_{2} = t \in \mathbb{Z}
\,.
\end{equation}

Following \eqref{eq: condition on Sigma for invariance of H - supplement}, 
we can parametrize the operator $\Sigma$ as
\begin{equation}
\Sigma(\alpha, \beta)
= 
e^{-\mathrm{i}\, \alpha \int^{x{R}}_{x_{L}} dx \, \partial_{x}(n \phi_{1} - \phi_{2}) }
\,
e^{\mathrm{i}\, (\beta/n) \int^{x{R}}_{x_{L}} dx \, \partial_{x}\phi_{2} }
\equiv
\tilde{S}_{\alpha}\,S_{\beta}
\,,
\quad
\alpha \in \mathbb{R} \,,~  \beta \in \mathbb{Z}
\,.
\end{equation}

$\tilde{S}_{\alpha}$ can be interpreted as a U(1) symmetry generator when the 
edge modes carry electromagnetic charge with the charge vector $t = (n,1)$:
\begin{subequations}
\label{eq: transformations of psi under Salpha and Sbeta}
\begin{equation}
\tilde{S}_{\alpha} : \psi_{1} \rightarrow \psi_{1} e^{\mathrm{i} 2 \pi n \alpha}\,,~ 
\psi_{2} \rightarrow \psi_{2} e^{\mathrm{i} 2 \pi \alpha}
\,, \quad \alpha \in \mathbb{R}
\,.
\end{equation}

The operator $S_{\beta}$, on the other hand, reflects the fact that the 
edge Hamiltonian gives rise to an $n$-particle condensate of phase 2;
associated to this interaction is the invariance
\begin{equation} 
S_{\beta} : \psi_{1} \rightarrow \psi_{1} \,,\psi_{2} \rightarrow \psi_{2} e^{\mathrm{i} 2 \pi \beta/n}
\,, \quad \beta \in \mathbb{Z}
\,.
\end{equation}
\end{subequations}

Moreover we find from \eqref{eq: transformations of psi under Salpha and Sbeta} that
\begin{equation}
\tilde{S}^{\dagger}_{\alpha}\,\sigma(x)\,\tilde{S}_{\alpha} = \sigma(x)
\,,
\quad
S^{\dagger}_{\beta}\,\sigma(x)\,S_{\beta} = \omega^{-\beta}\,\sigma(x)
\,.
\end{equation}

Of particular interest are the operators
\begin{equation}
\label{eq: definition of S - supplement}
\mathcal{S}
=
e^{-\frac{\mathrm{i}}{n} \int^{x{R}}_{x_{L}} dx \, \partial_{x}\phi_{2} } 
\,
\end{equation}
and
\begin{equation}
\label{eq: definition of A - supplement}
\mathcal{A}
=
e^{\mathrm{i}\, \int^{x{R}}_{x_{L}} dx \, \partial_{x}(n \phi_{1} - \phi_{2}) }
\,
e^{-\frac{\mathrm{i}}{n} \int^{x{R}}_{x_{L}} dx \, \partial_{x}\phi_{2} } 
=
\sigma^{\dagger}(x_{L}) \sigma(x_{R}) \, e^{-\frac{\mathrm{i}}{n} \int^{x{R}}_{x_{L}} dx \, \partial_{x}\phi_{2} } 
\equiv
\alpha^{\dagger}(x_{L})\,\alpha(x_{R})
\,,
\end{equation}
where 
$
\alpha(x_{L}) = \sigma(x_{L}) 
$
and
$
\alpha(x_{R})
=
\sigma(x_{R})\,e^{-\frac{\mathrm{i}}{n} \int^{x{R}}_{x_{L}} dx \, \partial_{x}\phi_{2} } 
$
are parafermion operators\cite{Fradkin-1980-suppl} defined at the ends of the interface. 
The topological property of the gapped interface is manifested by the presence of
these zero modes whereby the $n$-fold degeneracy is encoded in the $n$ eigenvalues of the 
operator $\mathcal{A}$, which commutes with $H_{int}$. 
Furthermore, after mapping parafermion operators into clock operators $ ( \sigma(x), \tau(x) )$,
the ground state $n$-fold degeneracy  becomes associated with the symmetry
breaking of the clock model, where the generator of the global $\mathbb{Z}_{n}$ symmetry
is $\mathcal{S} = \prod_{x} \tau(x)$. In the following we discuss this mapping
between clock and parafermion operators in further detail.

\subsection{Lattice Regularization:  Parafermion/Clock Representations}
\label{subsec: Lattice Regularization:  Parafermion/Clock Representations}

We now discuss a lattice regularization of the 
interaction Hamiltonian \eqref{eq: strongly coupled Hamiltonian - supplement}
responsible for gapping the interface between phases 1 and 2.
Such regularization serves a useful purpose in allowing a microscopic representation of the parafermion end states,
in connection with recent microscopic models of topological 
parafermion chains~\cite{Kitaev-2001-suppl,Fendley-2012-suppl,Bondesan-2013-suppl,Motruk-2013-suppl,Jermyn-2014-suppl,Zhuang-2015-suppl,Alexandradinata-2015-suppl}.

We address first the lattice regularization in terms of clock variables. 
As we have seen, the Hamiltonian \eqref{eq: strongly coupled Hamiltonian - supplement}
commutes with the operator $\mathcal{S}$ defined in Eq.~\ref{eq: definition of S - supplement}.
This operator can be identified with the generator of $\mathbb{Z}_{n}$
transformations of the clock operator $\sigma(x)$ via $\mathcal{S}^{\dagger}\,\sigma(x)\,\mathcal{S} = \omega\,\sigma(x)$.
In this language, the ground state degeneracy of the Hamiltonian \eqref{eq: strongly coupled Hamiltonian - supplement}
becomes associated with the ``symmetry broken" states
\begin{equation}
\label{eq: eigenstates of sigma - supplement}
\forall x:
~~
\sigma(x)\,\ket{\Psi_{q}}
=
\omega^{q}\,\ket{\Psi_{q}}
\,,
\quad
 q = 1, ..., n
\,,
\end{equation}
where $\omega \equiv e^{\mathrm{i}\frac{2\pi}{n}}$.
While it would seem that a particular eigenstate could be identified by the local measurement
of $\sigma(x)$, $\bra{\Psi_{q}} \sigma(x) \ket{\Psi_{q'}} = \omega^{q}\,\delta_{q,q'}$,
it is fundamental to recognize that $\sigma(x)$ is \textit{not} a local operator in terms of the 
original local degrees of freedom $\psi_{1}$ and $\psi_{2}$.

Despite its non-local character, it is still useful to identify $\sigma(x)$ as a $\mathbb{Z}_{n}$
clock operator, with $\sigma^{\dagger}(x) = \sigma^{(n-1)}(x)$, while acting on the ground state manifold
Eq.~\ref{eq: eigenstates of sigma - supplement}.
Note that, since $\Theta(x)$ is a bosonic operator, i.e., $[ \Theta(x), \Theta(x') ] $ = 0 
for any coordinates $x$ and $x'$, so is $\sigma$: $\sigma(x) \sigma(x') = \sigma(x') \sigma(x)$, for $x \neq x'$.
With that we are led to re-expresses the interaction \eqref{eq: strongly coupled Hamiltonian - supplement} 
using the relation
\begin{equation}
\label{eq: rotor model connection - supplement}
\begin{split}
e^{\mathrm{i} n\Theta(x)}
=
\textrm{lim}_{\delta \rightarrow 0}
\,
e^{\mathrm{i} \Theta(x)}
\,
e^{\mathrm{i} (n-1) \Theta(x + \delta)}
=
\textrm{lim}_{\delta \rightarrow 0}
\,
\sigma(x) \sigma^{\dagger}(x+\delta)
\,.
\end{split}
\end{equation}

Eq.\eqref{eq: rotor model connection - supplement} 
motivates introducing a lattice regularization 
by defining clock operators $\sigma_{i}$ at every site $i \in {1, ... , L}$ of the open chain and the canonically
conjugated operators $\tau_{i}$ satisfying $\tau^{\dagger}_{i} \sigma_{i} \tau_{i} = \omega \sigma_{i}$.
Then a lattice regularization of the interaction \eqref{eq: strongly coupled Hamiltonian - supplement} reads
$
H_{Lattice}
=
-J_{L}\,\sum^{L}_{i=1}
\sigma_{i} \sigma^{\dagger}_{i+1}
+
\textrm{H.c.}
$,
whose $n$ degenerate ground states satisfy
$
\sigma_{i} \ket{\tilde{\Psi}_{q}} = \omega^{q} \ket{\tilde{\Psi}_{q}}
$,
for $q = 1, ..., n$,
in direct correspondence with \eqref{eq: eigenstates of sigma - supplement}.

In the lattice representation, one introduces operators ($\alpha_{2i-1}, \alpha_{2i}$)
at every site of the open chain defined by~\cite{Fradkin-1980-suppl,Fendley-2012-suppl}
\begin{equation}
\label{eq: parafermions def lattice - supplement}
\alpha_{2j-1}
=
\sigma_{j}
\prod^{j-1}_{k=1} \tau_{k}
\,,
\quad
\alpha_{2j}
=
\omega^{(N-1)/2} \sigma_{j} \prod^{j}_{k=1} \tau_{k}
\,.
\end{equation}
These new operators satisfy parafermion statistics~\cite{Fradkin-1980-suppl}
\begin{equation}
\alpha_{i} \alpha_{j} = \omega^{\textrm{sgn}(j - i)} \alpha_{j} \alpha_{i}
\,.
\end{equation}

The field theory equivalent of the parafermion operators \eqref{eq: parafermions def lattice - supplement} reads
\begin{equation}
\label{eq: def parafermion - supplement}
\alpha(x)
=
\sigma(x)
\,
e^
{
-\frac{\mathrm{i}}{n}\,\int^{x}_{x_{L}} dz \, \partial_{z} \phi_{2}(z)		
}
\,,
\end{equation}
satisfying the commutation relations
\begin{equation}
\label{eq: parafermion algebra continuum - supplement}
\alpha(x)\,\alpha(y)
=
\alpha(y)\,\alpha(x)
\,
e^{\mathrm{i}\frac{2 \pi}{n} \textrm{sgn}(y - x)}
\,.
\end{equation}

In the parafermion representation the lattice Hamiltonian acquires
the form
\begin{equation}
\label{eq: lattice parafermion Hamiltonian - supplement}
H_{Lattice}
=
-J_{L}\,\sum^{L}_{i=1}
\omega^{(n-1)/2} \alpha^{\dagger}_{2i+1} \alpha_{2i}
+
\textrm{H.c.}
\,.
\end{equation}

Notably, $\alpha_{1}$ and $\alpha_{2L}$ do not appear in the Hamiltonian; for $n=2$ these ``dangling" parafermions
reduce to the Majorana end states in the topological Kitaev chain.~\cite{Kitaev-2001-suppl}
Manifestly, $[ H_{Lattice},~  \mathcal{A}_{lattice} ] = 0$, where 
$
\mathcal{A}_{lattice} = \alpha^{\dagger}_{1} \alpha_{2L}
$
is the non-local operator connecting the parafermion end state through the
gapped bulk.

\begin{figure}[h!]
\includegraphics[width=0.5\textwidth]{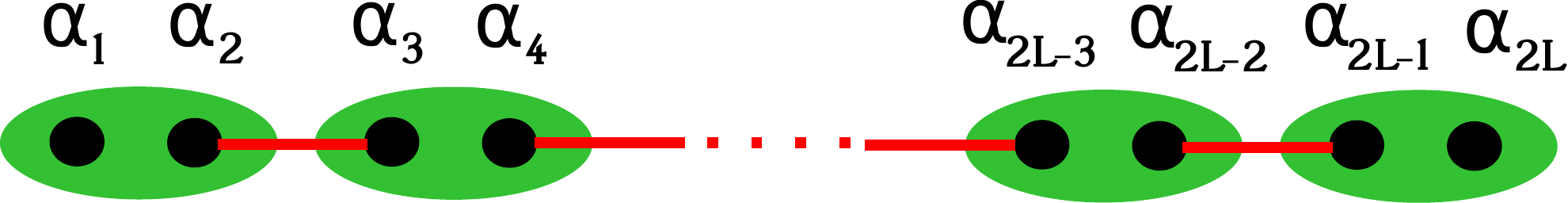}
\caption
{
(color online)
Lattice regularization of the Hamiltonian \eqref{eq: strongly coupled Hamiltonian - supplement}.
The non-trivial dimerization pattern encodes the presence of
parafermion excitations at the end points the of chain.
}
\label{parafermion-wire}
\end{figure}

\section{$2+1$D chiral topological state coupled to $\mathbb{Z}_{n}$ gauge theory}
\label{sec: topological state coupled to Zn gauge theory}

We now provide further details of the how the bulk phases 1 and 2 can be related to each other
by the confinement (or deconfinement) of a 2D $\mathbb{Z}_{n}$ gauge theory.  	
To start, let  phase 2  be coupled to a $\mathbb{Z}_{n}$ gauge theory in its deconfined phase
such that the gauge field $\alpha_{\mu}$ describes the excitations of phase 2, and ($a_{\mu},b_{\mu}$)
the excitations of the $\mathbb{Z}_{n}$ gauge theory. Hence, the coupled system is described by the Abelian Chern-Simons
theory:

\begin{subequations}
\label{eq: 2D topological phase coupled to Zn gauge theory - supplement}
\begin{equation}
\begin{split}
\mathcal{L}
=
\frac{1}{4\pi}\,\varepsilon^{\mu\nu\lambda}\,
c^{I}_{\mu}
\,
\bar{K}_{I J}(p,n)
\,
\partial_{\nu}\,c^{J}_{\lambda}
\,,
\end{split}•
\end{equation}•
where
\begin{equation}
\bar{K}(p,n)
=
\begin{pmatrix}
p & -1 & 0
\\
-1 & 0  & n
\\
0 & n & 0
\end{pmatrix}•
\,\quad
c
=
\begin{pmatrix}
\alpha
\\
a
\\
b
\end{pmatrix}•
\,.
\end{equation}•
\end{subequations}•
$
\bar{K}(2q+1, n)
$
and
$
\bar{K}(2q, n)
$
describe, respectively a $\nu = 1/(2q+1)$ fermionic and a $\nu = 1/(2q)$ bosonic Laughlin state coupled to a deconfined $\mathbb{Z}_{n}$ gauge theory.
The original charge and flux excitations of the $\mathbb{Z}_{n}$ gauge theory
are described, respectively, by $e = (0,1,0)$ and $m = (0,0,1).$

For the choice $p = 2q+1$, the GL(3, $\mathbb{Z}$) transformation
\begin{equation}
W_{F}
=
\begin{pmatrix}
n & 1 & -1
\\
n(2q+1) & q & -(q+1)
\\
1 & 0 & 0
\end{pmatrix}•
\,,\quad
\textrm{Det}[W_{F}] = -1
\,,
\end{equation}•
implements
\begin{equation}
\label{eq: transformed K-matrix of fermionic Laughlin state coupled to Zn gauge theory}
W^{T}_{F}
\,
\bar{K}(2q+1,n)
\,
W_{F}
=
(2q+1)\,n^{2}\,\oplus \sigma_{z}
\equiv
K_{g}(2q+1,n)
\,.
\end{equation}•

For the choice $p = 2q$, the GL(3, $\mathbb{Z}$) transformation
\begin{equation}
W_{B}
=
\begin{pmatrix}
-n & -1 & 0
\\
-2 q n & -q & 1
\\
-1 & 0 & 0
\end{pmatrix}•
\,,\quad
\textrm{Det}[W_{B}] = 1
\,,
\end{equation}•
implements
\begin{equation}
\label{eq: transformed K-matrix of bosonic Laughlin state coupled to Zn gauge theory}
W^{T}_{B}
\,
\bar{K}(2q,n)
\,
W_{B}
=
2q\,n^{2}\,\oplus \sigma_{x}
\equiv
K_{g}(2q,n)
\,.
\end{equation}

The edge states associated with the bulk theory \eqref{eq: 2D topological phase coupled to Zn gauge theory - supplement}
are described by the $1+1$D theory
\begin{equation}
\mathcal{L}
=
\frac{1}{4\pi}\partial_{t}\bar{\Phi}^{T}\,\bar{K}\,\partial_{x}\bar{\Phi}  + ....
=
\frac{1}{4\pi}\partial_{t}\Phi^{T}\,K_{g}\,\partial_{x}\Phi  + ....
\,,
\end{equation}
where $\bar{\Phi} = W\,\Phi$ sets the relation between the edge fields 
$\bar{\Phi}^{T} = (\bar{\phi}_{1},\bar{\phi}_{2},\bar{\phi}_{3})$ and
$\Phi^{T} = (\phi_1, \phi_2, \phi_3)$ of the two representations.

As discussed in the main text, in the representation given by $K_{g}$, 
the interaction for the case $p=2q+1$
\begin{equation}
\label{eq: interaction V1 - supplement}
V_{1}
=
\cos
{
\left(
L_{1}^{T}\,K_{g}\,\Phi
\right)
}
\,,
\quad
L^{T}_{1} = (0,1,1)
\,
\end{equation}
gaps the ``extra" trivial modes, which then yields the edge state of phase $1$.
Alternatively, the interaction
\begin{equation}
\label{eq: interaction V2 - supplement}
V_{2}
=
\cos
{
\left(
L_{2}^{T}\,K_{g}\,\Phi
\right)
}
\,,
\quad
L^{T}_{2} = (1,q\,n,(q+1)\,n)
\,
\end{equation}
gives rise to the edge state of phase 2.

We can now establish a relation between null vectors
in the two representations by noticing that, if $L$ is a null vector of $K_{g}$,
such that, $L^{T}\,K_{g}\,L = 0$, then $\bar{L} = W\,L$ is the corresponding null vector of $\bar{K}$, since
$\bar{L}^{T}\,\bar{K}\,\bar{L} = L^{T} W^{T} \bar{K} W L = L^{T}\,K_{g}\,L = 0$.
By making use of this relationship, we find
\begin{equation}
\label{eq: interaction V1 Kbar representation - supplement}
V_{1}
=
\cos
{
\left(
\bar{L}_{1}^{T}\,\bar{K}\,\bar{\Phi}
\right)
}
=
\cos
{
\left(
\bar{\phi}_1 - n \bar{\phi}_3
\right)
}
\,,
\quad
\bar{L}^{T}_{1}  = (W L_{1})^{T} = (0,-1,0)
\,
\end{equation}
and
\begin{equation}
\label{eq: interaction V2 Kbar representation - supplement}
V_{2}
=
\cos
{
\left(
\bar{L}_{2}^{T}\,\bar{K}\,\bar{\Phi}
\right)
}
=
\cos
{
\left(
n \bar{\phi}_2
\right)
}
\,,
\quad
\bar{L}^{T}_{2}  = (W L_{2})^{T} = (0,0,1)
\,.
\end{equation}
It is clear then from Eqs.~\eqref{eq: interaction V1 Kbar representation - supplement}
and~\eqref{eq: interaction V2 Kbar representation - supplement} that the
vacuum associated with $\langle\,V_2\,\rangle \neq 0$ confines the excitations of the $\mathbb{Z}_n$ gauge theory yielding the original phase 2.
Moreover, in the basis given by $\bar{K}$, the deconfined excitations in the vacuum
where $\langle\,V_1\,\rangle \neq 0$ are described by vertex operators $\exp{\left( \bar{\ell}^{T} \bar{\Phi} \right)}$,
with $\bar{\ell}^{T} = (\bar{\ell}_1, \bar{\ell}_2, \bar{\ell}_3)$, such that 
$\bar{\ell}^{T} \bar{L}_{1} = -\bar{\ell}_2 = 0$. Then computing the braiding statistics of two 
excitations $\bar{\ell} = (\ell_1,0,\ell_3)$ and $\bar{m} = (m_1,0,m_3)$ we find
$\ell^{T}\,\bar{K}^{-1}\,\bar{m} = \frac{(\ell_3 + n \ell_1)(m_3 + n m_1)}{(2q+1) n^{2}}$,
which accounts for the chiral phase with $k_1 = (2q+1) n^2$.
This result is consistent with the topological state
having $p n^2$ classes of quasiparticles $\{ m^{1}, \ldots , m^{p n^2} \}$,
where $m^{1}$ represents the deconfined magnetic flux originating from the $\mathbb{Z}_n$ gauge
theory and $\psi_1 = m^{p n}$ represents the fermion (boson) of the $k = p$ theory [for $p$ even (odd)],
which becomes a non-local quasiparticle upon coupling to the deconfined gauge theory.

\section{Gapped parafermion interface behaves as an anyon Andreev reflector}
\label{sec: Gapped parafermion interface behaves as an anyon Andreev reflector}
Setting $p=1$ for simplicity, suppose a gapped interface between phases 1 and 2 is formed,
which physically implies the condensation of the bosonic field
$\Theta(x) = n \phi_{1}(x) - \phi_{2}(x)$, so that
$
\langle\,e^{\mathrm{i} \Theta} \, \rangle = 
\langle\,e^{\mathrm{i} (n \phi_{1} - \phi_{2})} \, \rangle 
\propto
\langle\,\chi_{1} \bar{\psi}_{2} \, \rangle 
\neq 0
$.
(Equivalently, $\langle\,e^{-\mathrm{i} \Theta} \, \rangle \propto \langle\,\bar{\chi}_{1} \psi_{2} \, \rangle \neq 0$.)
Hence, the gapped spectrum originates from the condensation of $\chi_{1} \bar{\psi}_{2}$, or,
equivalently, of $\bar{\chi}_{1} \psi_{2}$.
This anyon condensate implies that the gapped interface behaves as a anyon ``Andreev reflector" in the following sense.

Let an excitation $\psi_{2}$ approach the gapped edge (for $p=1$, this is the only
quasiparticle of the phase 2). Moreover, suppose that the pair $\chi_{1} \bar{\chi}_{1}$
is produced from vacuum fluctuations on the topological system 1 near the interface.
Then due to the nature of the condensation on the interface, the pair $\bar{\chi}_{1} \psi_{2}$ can be condensed
and absorbed by the interface, while leaving behind the deconfined excitation $\chi_{1}$ on the other side of the interface. 

Reversing the logic, we can ask: what happens if a quasiparticle $\varepsilon^{\ell}_{1}$ ($\ell \in \{ 1, ... , n^{2} \}$)
belonging to phase 1 approaches the interface where it interacts with the condensate?
Since the only allowed condensation process involves the quasiparticle $\chi_{1}$,
the following non-trivial process can occur:
close to the interface, (1) the incoming quasiparticle decays into 
$\varepsilon^{\ell}_{1} = \varepsilon^{\ell - p n}_{1}\,\chi_{1}$ and, 
(2) the pair $\psi_{2} \bar{\psi}_{2}$ is produced from vacuum fluctuations on the topological system 2 
near the interface such that, when the condensation $\langle \chi_{1} \bar{\psi}_{2} \rangle \neq 0$ occurs, it leaves
behind the deconfined excitations $\varepsilon^{\ell - p n}_{1} = \varepsilon^{\ell}_{1} \bar{\chi_{1}}$
and $\psi_{2}$ on different sides of the interface.
In particular, the case $\ell = p n$ represents an incoming $\chi_{1}$ quasiparticle,
which can be completely absorbed by the interface while the $\psi_{2}$
excitation appears on the other side.

\section{Commutation relation of vertex operators}
\label{sec: Commutation relation of vertex operators}

Define the operators:
\begin{equation}
\label{eq: definition operators gamma}
\begin{split}
\Gamma_{k}
=
e^
{
\mathrm{i}\,
c_{k}\,
\int^{z_{k}}_{y_{k}}\, dx\,
\partial_{x}\,
\left(
L^{T}_{k} \cdot \Phi(x)
\right)
}
=
e^
{
\mathrm{i}\,
c_{k}\,
\Big[
L^{T}_{k} \cdot \Phi(z_{k})
-
L^{T}_{k} \cdot \Phi(y_{k})
\Big]
}
\,,
\end{split}•
\end{equation}•
where 
$
c_{k} \in \mathbb{R}
$
is a real coefficient and 
$
(y_{k}, z_{k})
$
is a finite interval on the line.
It follows that
\begin{equation}
\label{eq: algebra operators gamma}
\begin{split}
&\,
\Gamma_{k}\,\Gamma_{p}
=
\Gamma_{p}\,\Gamma_{k}
\,
e^{\mathrm{i}\,\Theta_{k p}}
\,,
\\
&\,
\Theta_{k p}
=
- \pi\,c_{k}\,c_{p}
\,
L^{T}_{k} \cdot K^{-1} \cdot L_{p}
\,
\Big[
\textrm{sgn}(z_{k}-z_{p})
-
\textrm{sgn}(z_{k}-y_{p})
-
\textrm{sgn}(y_{k}-z_{p})
+
\textrm{sgn}(y_{k}-y_{p})
\Big]
\,.
\end{split}•
\end{equation}•
It can be easily seen from the commutation relations of the edge fields
that, if two intervals $(y_{k}, z_{k})$ and $(y_{p}, z_{p})$ are non overlapping
or one of intervals is entirely contained within the other,
than $\Theta_{k p} = 0$, which implies that $[ \Gamma_k, \Gamma_p] = 0$.
If, however, the intervals $(y_{k}, z_{k})$ and $(y_{p}, z_{p})$ are partially overlapping such that $y_k < y_p < z_k < z_p$, then it follows that 
$
\Theta_{k p}
=
2\,\pi\,c_k\,c_p\,
L^{T}_{k} \cdot K^{-1} \cdot L_{p}
$.

\end{document}